# The trigonal structure as a reference to access the spontaneous polarization of wurtzite crystals

Abdesamed Benbedra[a], Said Meskine[a], Abdelkader Boukortt[a], Roland Hayn[b]*, Hamza Abbassa[a], and El Habib Abbes[a]

[a] Laboratoire d'Elaboration et Caractérisation Physico Mécanique et Métallurgique des Matériaux (ECP3M), Université Abdelhamid Ibn Badis, Route Nationale N°11, Kharouba, 27000 Mostaganem, Algérie
[b] Aix-Marseille Université, CNRS, IM2NP-UMR 7334, 13397 Marseille Cedex 20, France
*Corresponding author: roland.hayn@im2np.fr

**Abstract**

The spontaneous polarization of wurtzite III-V nitrides XN (X=Al, Ga, In) and II-VI oxides YO (Y=Be, Zn) is investigated via first-principles computational methods. The modern treatment defines this quantity as the polarization difference between the investigated system and an appropriate reference state. We demonstrate that the trigonal structure can be used as a reference to determine the spontaneous polarization of wurtzite materials. We compare the current values with the widely-known zincblende results reported in the literature and find a very good agreement. It is shown that the electronic contribution of polarization is greater than the ionic one. Furthermore, we reproduce the experimental value of the spontaneous polarization of wurtzite GaN reported in a previous study. In order to do so, we calculate the spontaneous polarization for each type of stacking faults using periodic supercells and the Berry-phase method. This theoretical analysis leads to a value nearly identical to the experimental measurement.

## I. Introduction

Due to their intrinsic low-symmetry, wurtzite crystals are characterized by the existence of a built-in electric polarization, called spontaneous polarization, which persists even in equilibrium [1]. The spontaneous polarization manifests itself as bound charges and internal electric fields in the layers of heterostructures, significantly affecting the properties and performance of semiconductor devices, such as light-emitting diodes, lasers, and high mobility transistors [2,3]. Experimental measurements of the spontaneous polarization are very limited, simply because such measurements are too challenging [4]. Part of this difficulty is due to the fact that wurtzite crystals are pyroelectrics: the orientation of their polarization is fixed and cannot be reversed [4]. In ferroelectrics, on the other hand, application of a sufficiently intense electric field inverts the polarization vector. This polarization-reversal effect allows an accurate measurement of the spontaneous polarization and is the basis of many important applications [5]. Except for GaN [6], no direct determination for the spontaneous polarization is available, and the experimental values for the majority of wurtzite materials are still unknown. Because of this, theoretical calculations are currently used to provide polarization values in order to interpret the results of experiments and for device modeling.

The theoretical framework in which the polarization of solids is studied is the Berry-phase method, developed in the 90s by Resta, Vanderbilt and King-Smith [7,8]. This method is based on two fundamental insights. First, only polarization differences between two states are accessible and rigorously defined, experimentally as well as theoretically. Second, the difference in polarization is directly related to a quantum geometric phase of the electronic wavefunctions known as the Berry phase [9]. In the specific case of the spontaneous polarization, the two states correspond to a noncentrosymmetric structure (wurtzite in the present case) and its centrosymmetric counterpart, also referred to as the reference structure [10,11]. For materials crystallizing in the wurtzite phase, the cubic zincblende structure has been used in the literature as a reference to calculate the spontaneous polarization [12-16]. In this paper, we propose the trigonal structure as another reference in terms of which the spontaneous polarization of wurtzite crystals is defined. We show that the values that result using this reference are very close to those obtained by zincblende. Furthermore, the trigonal structure can be easily extended to include the effect of stacking faults on the spontaneous polarization. Because of their importance in modern technology, the binary III-V nitrides and II-VI oxides are chosen as case-studies to elucidate the equivalence between the two reference structures in determining the spontaneous polarization of wurtzite compounds.

As mentioned above, the spontaneous polarization of GaN was directly obtained from experiment [6]. This was achieved by employing the energy of excitons bound to the stacking faults in the crystal. Here, we reproduce the reported value using first-principles calculations. Specifically, we compute the spontaneous polarization associated with different types of stacking faults in wurtzite GaN. As will be shown later on, the spontaneous polarization of the perfect (defectless) crystal is much greater than that in the presence of stacking faults. This discrepancy is one of the issues that will be addressed in the present work. The rest of the paper is organized as follows. In section II, we describe both wurtzite and trigonal structures, and provide the numerical details of the simulation. Section III presents the results of the paper: section III.1 deals with the structural optimization needed to perform the calculations. Section III.2 gives the spontaneous polarization values referenced to the trigonal structure. In section III.3 we discuss the electronic and ionic contributions. We present the results of the spontaneous of stacking faults in GaN and explain the procedure carried out to determine them in section III.4. Finally, in section IV a conclusion is drawn which summarizes our main results.

## II. Theoretical approach

### II.1 Wurtzite and trigonal structures

The wurtzite structure has a hexagonal Bravais lattice (space group P6$_3$mc or 186) with four atoms per unit cell and two lattice constants: the edge length *a* of the hexagon base and the height *c* of the hexagonal prism. However, there is a third structural degree of freedom, a dimensionless internal parameter *u* that characterizes the cation-anion bond length parallel to the *c*-axis [14]. A schematic of the wurtzite unit cell is illustrated in figure 1 (*a*). The cations occupy (1/3,2/3,0) and (2/3,1/3,1/2) positions, while the anions are located at (1/3,2/3,*u*) and (2/3,1/3,*u*+1/2). The two types of atoms are tetrahedrally coordinated: each atom is surrounded by four atoms of the other type situated at the corners of a tetrahedron and vice versa. The three bonds in the basal plane are equivalent, but the bond along the *c*-axis is a little bit longer. This makes the wurtzite structure noncentrosymmetric, i.e. it lacks a center of inversion. This symmetry reduction, in addition to the ionic nature of the bonding, are the physical origin of the spontaneous polarization [17].

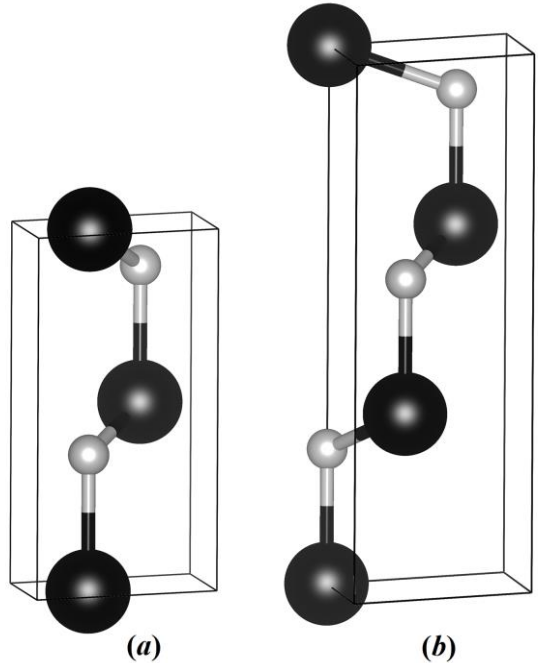

**Figure 1** Unit cell of the (*a*) wurtzite and (*b*) trigonal structures. Both structures are tetrahedrally bonded, but they differ in the stacking sequence of atomic planes: AB for wurtzite and ABC for trigonal. Black and grey spheres correspond to cations and inions, respectively.

Figure 1 (*b*) shows the unit cell of the trigonal structure (P3m1 or 156). It has a hexagonal-like unit cell containing six atoms. The positions of the cations are (0,0,0), (1/3,2/3,1/3) and (2/3,1/3,2/3), and those of the anions are (0,0,1/4), (1/3,2/3,7/12) and (2/3,1/3,11/12). In such a way, the atomic positions are identical to zincblende using a different space group: (i) the bonding is tetrahedral with the four bonds having the same length, and (ii) the stacking of atomic planes is in the sequence ABC along the [111] direction, which is in opposition to the wurtzite system, where the stacking is AB along [0001]. Because of these similarities, the trigonal structure is sometimes referred to as the six-atom [111]-oriented zincblende structure [18]. Our major finding is that the trigonal structure is well suited for the derivation of the spontaneous polarization of wurtzite materials, yielding similar results as the ordinary zincblende phase.

## II.2 Computational details

The first-principles results presented in this work are obtained using density functional theory (DFT) [19] within the framework of the Kohn-Sham (KS) algorithm [20] as incorporated in the computer code WIEN2k [21]. In all calculations, the exchange-correlation contribution to the total energy is described by the combination of the general gradient approximation [22] and the modified Becke-Johnson approach [23,24] (GGA+mBJ). To solve the KS equations we rely on the full-potential linearized augmented plane-wave (FP-LAPW) method [21,25], which consists of splitting the unit cell using a spherical boundary (called muffin-tin sphere) around the atomic cores. This allows us to expand the solutions on different basis sets inside and outside the spheres: nonlocalized plane waves on the outside, and localized atomic-like orbitals inside. In principle, these expansions run to infinity, but in a real calculation they have to be limited at some point, this is called the expansions cutoff. In our study the development on atomic orbitals is up to a maximum angular momentum $l_{max}$=10, and a cutoff of $R_{MT} K_{max}$=8.5 is used for the plane-wave expansion, where $R_{MT}$ is the radius of smallest muffin-tin sphere (see below), and $K_{max}$ represents the norm of largest wave vector.

For energy calculations, we need to solve the Schrödinger-like KS equations everywhere in the first Brillouin zone (BZ) of the crystal. In practice, we do it for a finite number of *k*-values, and get a value for the energy at each *k*. The tetrahedron method is employed to perform the integration over the BZ by taking a dense mesh of 1000 *k*-points (a uniform sampling of 10×10×10). These values of the expansion cutoffs and *k*-points are found to be enough for the convergence of the total energy and electric charge to an accuracy of about $10^{-4}$ Ry and $10^{-3}$ *e*, respectively.

FP-LAPW is an all-electrons technique, meaning it considers all electrons self consistently in a full-potential treatment [21,25]. The core electrons are fully confined within the muffin-tin spheres while valence electrons are not. The radii $R_{MT}$ of these atomic spheres for different elements are chosen as follows: 1.5, 1.8, 2, 2.1, 1.6, 1.4, and 2.05 Bohr for N, Al, Ga, In, O, Be and Zn, respectively. We take the borderline to distinguish between core and valence electrons to be −6 Ry, i.e. all orbitals with an energy more negative than this value are core states.

The computation of polarization properties is carried out via the code BerryPI [26], which allows to calculate the electronic and the ionic contributions. The electronic part of the spontaneous polarization is evaluated as a Berry phase of the cell-periodic Bloch functions. For the sake of consistency and maintaining the same degree of convergence, we kept the number of *k*-points in BerryPI the same as in total-energy calculations.

### III. Results and discussion

### III.1 Structural optimization

In order to determine the spontaneous polarization for each material, two structures must be carefully prepared. These are the wurtzite and the reference trigonal structures. For the wurtzite system, the determination of structural properties is performed by means of total-energy minimization process [27]. The equilibrium lattice constants *a* and *c* are obtained from a fit of the system energy as a function of unit cell volume and *c*/*a*-ratio. The internal parameter *u* is obtained by the relaxation of atomic positions according to the Hellman-Feynman theorem [28], such that the convergence criterion on the forces is 0.1 mRy/Bohr. The results of the structural optimization are shown in table 1 for each wurtzite compound. We find an overall agreement between our calculated values and those obtained by experiment [29-31]. The values of lattice constants *a* and *c* obtained by GGA+mBJ tend to be higher than those obtained experimentally. The relative differences are only about 1%, which is typical for well-converged DFT-calculations. The theoretical internal parameters agree very well with the experimental ones. For all wurtzite crystals, $u$ is slightly greater than the ideal value 0.375.

**Table 1** Calculated equilibrium lattice constants $a$ and $c$ (in Å) and the internal parameter $u$ (in units of $c$) within the GGA+mBJ approximation for wurtzite AlN, GaN, InN, BeO and ZnO.

| Material | Wurtzite | | | Trigonal | | |
|---|---|---|---|---|---|---|
| | $a$ | $c$ | $u$ | $a$ | $c$ | $u$ |
| AlN | 3.13 | 5.02 | 0.381 | 3.13 | 7.67 | - |
| GaN | 3.23 | 5.25 | 0.376 | 3.23 | 7.91 | - |
| InN | 3.59 | 5.79 | 0.379 | 3.59 | 8.79 | - |
| BeO | 2.72 | 4.41 | 0.378 | 2.72 | 6.65 | - |
| ZnO | 3.29 | 5.28 | 0.381 | 3.29 | 8.06 | - |

It is not necessary to fully optimize the geometry of the trigonal structure, because its in-plane lattice parameter $a$ is required to be the same as that of the wurtzite phase [16], such that the mismatch $\Delta a/a$ is identically zero. As for the out-of-plane lattice constant $c$, it is obtained from the lattice-constant ration $c/a$ as follows. The ideal hexagonal $c/a$ equals $\sqrt{8/3}$ [32]. Since the trigonal system has three atomic layers [see Fig. 1 (*b*)], and the thickness of one layer is $0.5c$ [33], the trigonal ratio $c/a$ is simply $1.5\sqrt{8/3}$. With a bit of algebra, the lattice constant $c$ is given by $c=a\sqrt{6}$. We report in table 1 the values of the trigonal lattice parameters. Note that unlike the wurtzite phase the trigonal one does not contain internal degrees of freedom.

### III.2 Spontaneous polarization

Once the structures are prepared, we perform a self-consistent field (SCF) calculation to determine the wavefunctions, from which the spontaneous polarization $P_{sp}$ is evaluated with the Berry-phase technique. This is done, as outlined in section I, by computing the difference between the polarization of the relaxed wurtzite and the trigonal structures:

$$P_{sp} = P_{WZ} - P_T, \qquad (1)$$

where WZ and T denote the wurtzite and trigonal polarizations, respectively. Within the Berry phase scheme, $P_{WZ}$ and $P_T$ are usually called the formal polarizations, and the difference $P_{WZ} - P_T$ is the effective polarization [10,11]. We stress that this difference, not the formal polarization by themselves, is the quantity of interest. In table 2 we list the formal polarization as well as the effective spontaneous polarization values for both the nitrides and oxides referenced to the trigonal structure, together with the results of previous calculations that are based on cubic zincblende for the sake of comparison.

**Table 2** Berry-phase results of electric polarization (in C/m²) for AlN, GaN, InN, BeO, and ZnO in the wurtzite phase. The spontaneous polarization $P_{sp}$ is the difference between the formal polarizations $P_{WZ}$ and $P_T$ as described by Eq. 1.

| Material | $P_{WZ}$ | $P_T$ | $P_{sp}$ | Other calc. of $P_{sp}$ |
|----------|----------|-------|----------|--------------------------|
| AlN | -0.558 | -0.471 | -0.087 | -0.090 [a] |
| GaN | -0.480 | -0.445 | -0.035 | -0.034 [a] |
| InN | -0.406 | -0.360 | -0.046 | -0.042 [a] |
| BeO | -1.299 | -1.255 | -0.044 | -0.036 [b], -0.045 [c] |
| ZnO | -0.893 | -0.850 | -0.043 | -0.057 [b], -0.032 [d] |

[a] Ref. [34], [b] Ref. [15], [c] Ref. [14], [d] Ref. [35]

It appears that wurtzite nitrides and oxides are highly polarized: their spontaneous polarization is quite large, only one order of magnitude smaller than that of typical ferroelectrics [26]. The spontaneous polarization has a negative sign and is nonzero only along the *c*-axis, which corresponds to the [000$\bar{1}$] direction. This is expected since the intrinsic asymmetry of the bonding is parallel to this specific direction. In the present theory, the polarization is only unambiguously defined if its value is much smaller than the polarization quantum $ec/V$ [11], where $c$ is the lattice constant in the direction of the polar axis and V is the volume of the wurtzite unit cell. The value of this quantum is found to be around 2 C/m² for all compounds, two orders of magnitude greater than the calculated values of the spontaneous polarization, indicating that our results are physically meaningful.

As can be readily seen from the data in table 2, our values derived using the trigonal structure are in good agreement with the values reported in the literature obtained with zincblende. This agreement holds very well for the nitrides. There is a moderate scatter in the theoretically determined spontaneous polarizations for the oxides. However, our values are within the range of those reported in the literature. We have therefore verified that the trigonal and zincblende reference structures provide almost similar values when the spontaneous polarization of wurtzite crystals is concerned.

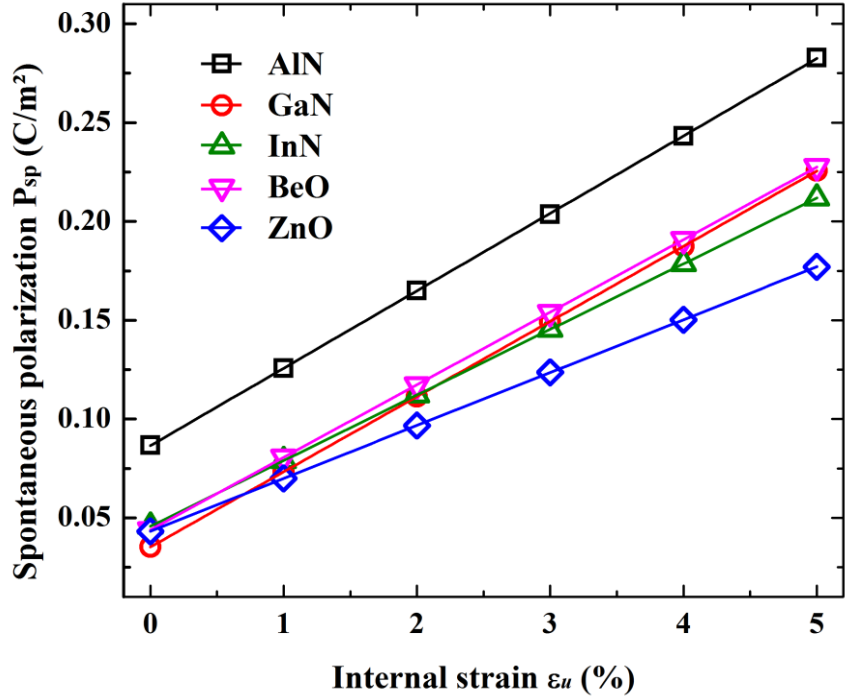

**Figure 2** Internal-strain dependence of the spontaneous polarization (in absolute value) for wurtzite nitrides and oxides. The values at zero strain correspond to those reported in table 2. The curves result from a linear regression of the calculated data.

As a final point, we investigate the linearity of the spontaneous polarization as a function of the internal parameter $u$. For this purpose, the spontaneous polarization is evaluated for internal strains $\varepsilon_u$ ranging from 0 to 5%. The internal strain is given by $\varepsilon_u = u - u_0/u_0$, where $u_0$ is the equilibrium internal parameter and $u > u_0$. The hexagonality of the structure is held constant throughout the calculation, since $u$ can take on different values without altering the symmetry of the unit cell. The results of this investigation are depicted in figure 4. The polarization is generally largest in AlN, smallest in ZnO, and intermediate in BeO, GaN and InN. The spontaneous polarization is observed to vary linearly and remains linear over the entire range of internal strain. The same result was previously obtained in the case of ferroelectric perovskites [36,37], which makes this kind of linearity a general feature of systems exhibiting a spontaneous polarization.

### III.3 Electronic and ionic contributions

Electric polarization can be decomposed into two distinct terms as follows [10,11]:

$$P_{sp} = P_{ele} + P_{ion}, \qquad (2)$$

where $P_{ele}$ and $P_{ion}$ are respectively the electronic and ionic contributions to the spontaneous polarization. The Berry-Phase method is only used to evaluate the electronic part of polarization, and has nothing to do with the determination of the ionic part. This is because electrons are described by quantum wavefunctions, while ionic cores are treated as classical point charges. To provide further insight into understanding the origin of the strong polarization in wurtzite crystals, we compute each term separately. The calculated contributions using the trigonal phase as a reference are reported in table 3.

**Table 3** Calculated electronic $P_{ele}$ and ionic $P_{ion}$ contributions (in C/m²) of the spontaneous polarization for each crystal studied, obtained using the trigonal reference structure.

| Material | AlN | GaN | InN | BeO | ZnO |
|---|---|---|---|---|---|
| $P_{ele}$ | -0.211 | -0.066 | -0.102 | -0.143 | -0.166 |
| $P_{ion}$ | 0.124 | 0.031 | 0.056 | 0.099 | 0.123 |

We observe that the two terms are of opposite sign. Specifically, the electronic part is negative while the ionic one is positive. This sign alternation seems typical for all tetrahedrally bonded materials. In addition, they have different magnitudes, with the electronic term being the largest, resulting in a negative value of the total polarization. One can safely anticipate, as remarked by Ahmed et *al.*, that in nonpolar crystals the absolute value of each of the terms is roughly the same, so that they tend to cancel each other, yielding zero net polarization [26]. In contrast, this cancellation is much less effective in wurtzite compounds, which is the reason of their large polarization.

**III.4 Spontaneous polarization of stacking faults in GaN**

**A. Stacking faults in wurtzite crystals**

The wurtzite structure in which GaN crystallizes is formed by the stacking of atoms in the sequence ..AB.. along the [0001] direction, where every layer is surrounded by two identical layers (e.g. ABA). In this case A-B bonds are called hexagonal bonds [38,39]. The zincblende (or trigonal) structure, on the other hand, can be described by the stacking sequence ..ABC.. along the [111] direction, in which the two layers surrounding a given layer are different (e.g. ABC), and the A-B bonds are called cubic bonds. Stacking faults can then be viewed as local deviations of the perfect wurtzite stacking sequence. In other words, stacking faults are embedded cubic units surrounded by a hexagonal matrix.

There are three types of basal-plane stacking faults reported in the literature: two intrinsic stacking faults of type-I and -II, and one extrinsic stacking fault. The type-I stacking fault originates from one violation of the wurtzite stacking rule, it introduces one cubic bond. The type-II stacking fault results from two violations of the normal stacking role and introduced two cubic bonds. Finally, the extrinsic stacking fault contains an additional C layer inserted in the midst of the original ..AB.. stacking sequence. This stacking fault introduces three cubic bonds [38,39].

In this work we reproduce the reported experimental value of the spontaneous polarization of GaN via ab-initio calculations, to be described shortly. However, we first present a brief overview of the experimental method used in Ref. 6. This method is based on a model in which the stacking faults are treated as a plate capacitor. The output of the measurement is the magnitude of the sheet charge that built-up at the wurtzite/zincblende interface, which is directly related to the spontaneous polarization of the low-symmetry structure:

$$P_{sp} = \varepsilon_0 \varepsilon \frac{\Delta V}{\Delta d}. \qquad (3)$$

In the above expression, $\Delta d$ is the difference in thickness between each of the stacking faults, $\varepsilon$ is the dielectric constant and $\varepsilon_0$ is the permittivity of free space. The potential change $\Delta V$ in this case is equal to the difference in transition energies between the stacking faults. The latter is the key ingredient of this model: the use of equation 1 for determining $P_{sp}$ requires at least a rough estimate of the transition energies of all types of stacking faults. This was done by two sets of spectroscopic measurements performed on samples of strain-free GaN microcrystal. Interested readers are referred to Refs. [6,33] for more details.

**B. Supercell structure**

In order to model the stacking faults, we use periodic supercells containing 20 atoms, i.e. 10 atomic layers, in a construction of 1×1×5 wurtzite unit cells. The different types of stacking faults are achieved by a proper manipulation of specific atomic positions applied to the original wurtzite stacking sequence as described above. The adopted supercells of the stacking faults in GaN are presented in figure 3. Due to imposing periodic boundary conditions, the beginning and the end of the supercells along the *c*-axis have to match. This has the effect of doubling the number of cubic bonds for each type of stacking faults. Indeed, the type-I stacking fault supercell shown in Fig. 3(*b*) contains two cubic bonds instead of one. The same remark can be made for type-II and extrinsic stacking fault supercells, in which there are four and six cubic bonds respectively (instead of two and three). For supercell calculations, which are more computationally expensive, the wave-plane expansion cutoff and the sampling in *k*-space reported in Sec. II.2 are reduced to 7 and 300, respectively.

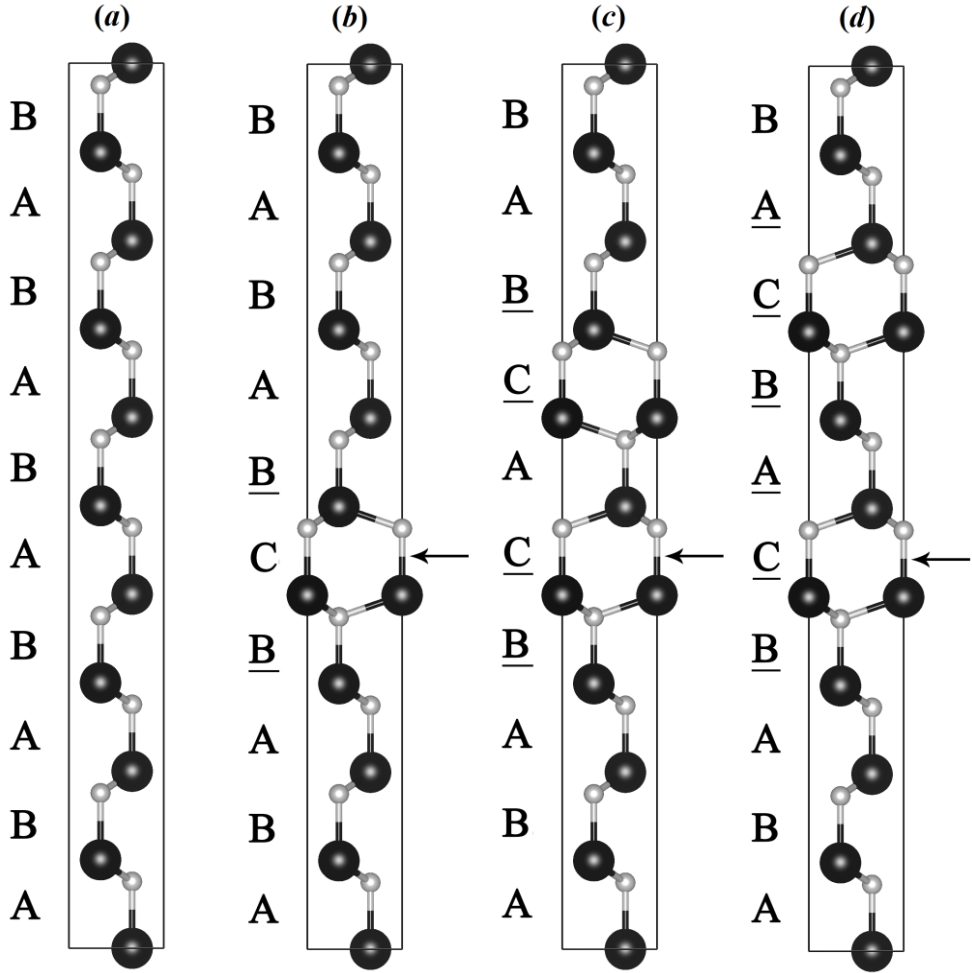

**Figure 3** Supercell models representing the studied structures for (*a*) ideal crystal, (*b*) type-I, (*c*) type-II, and (*d*) extrinsic stacking faults. The position of the stacking faults is indicated by an arrow, and the cubic bonds are indicated by underlined letters. Black and gray spheres correspond to Ga and N atoms, respectively.

## C. Results

We begin by determining the equilibrium structure of each stacking fault. We find that the presence of these defects has a negligible effect on the lattice parameters of wurtzite GaN reported in table 1, which means that the cubic units are almost perfectly lattice-matched to the hexagonal host. This result is in accordance with a previous study of stacking faults in nitrides [38]. Also, the atomic positions in the neighborhood of the stacking faults have been relaxed. Using the theoretical approach followed throughout this work, the spontaneous polarization of each type of stacking faults is calculated as the difference:

$$P_{sp}^{SF} = P_{WZ}^{SF} - P_T, \qquad (4)$$

where $P_{WZ}^{SF}$ is the formal polarization of the wurtzite supercell containing the stacking fault, and $P_T$ is the trigonal formal polarization. The calculated values for the intrinsic and extrinsic stacking faults are given in the histogram of figure 4. Also included is the $P_{sp}$ value of the ideal structure, where "ideal" refers to the supercell without stacking faults [Fig3. (*a*)]. It is important to observe the trend followed by the stacking faults: the spontaneous polarization decreases in magnitude with the number of cubic bonds, i.e. from type I to type II and, more significantly, to extrinsic stacking fault. Indeed, the extrinsic stacking fault has the closest structure to the nonpolar zincblende phase (with three cubic bonds); consequently, its polarization is the smallest among the stacking faults studied here. The supercell value of the perfect crystal is the same as the one obtained using an ordinary unit cell. This structure does not contain any cubic bonds by construction, this is why it has the maximum value of the spontaneous polarization.

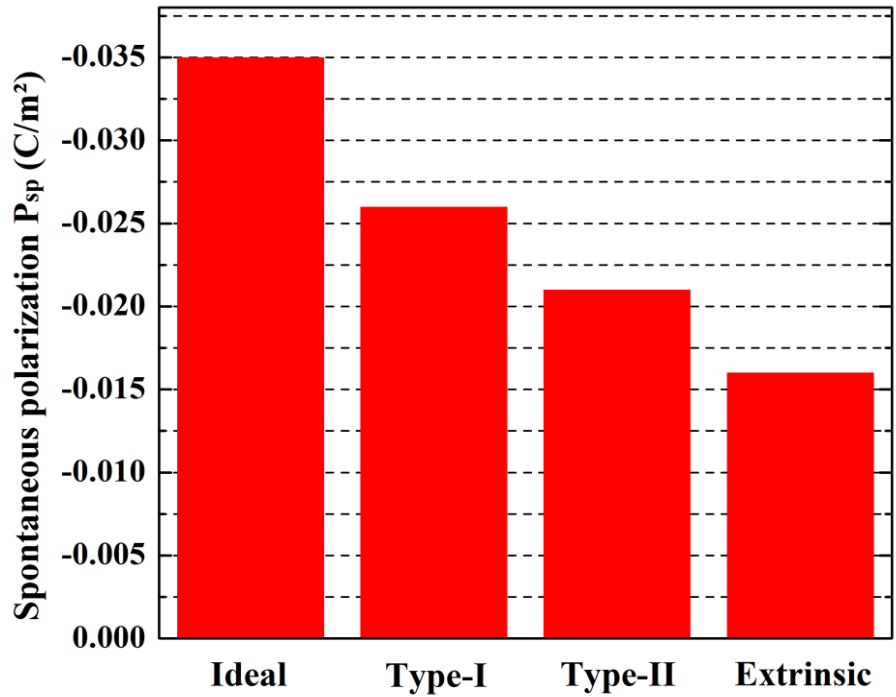

**Figure 4** Spontaneous polarization for the ideal and stacking faults structures of wurtzite GaN. All values of $P_{sp}$ are referenced to the trigonal structure.

To confirm the validity of the plate capacitor model, equation 3 was used in [6] to estimate the spontaneous polarization for the extrinsic stacking fault. In this case, $\Delta V$ and $\Delta d$ are obtained from self-consistent Poisson-Schrödinger calculations in Ref. [6]. This results in a value of -0.018 C/m², which is very close to our value of -0.016 C/m² for the same stacking fault (see Fig. 4).

The spectroscopically observed $\Delta V$ in Eq. 3 is actually the *average* value of the transition energy difference between the three types of stacking faults. So in a sense, the recommended experimental value of -0.022 C/m² is the spontaneous polarization of type I, type II and extrinsic stacking faults combined. Motivated by this, we compute the average of the three spontaneous polarization values given in figure 4. The value that we get is almost identical to the experimental one, namely -0.021 C/m². This discussion shows that assigning the experimental value of [6] or the average theoretical value reported here to GaN is somehow misleading: -0.022 C/m² or -0.021 C/m² is not the spontaneous polarization of the perfect defectless GaN, it is rather the value associated with the different types of stacking fault of this material.

## IV. Conclusion

In this paper, we have performed a theoretical study of the spontaneous polarization in wurtzite crystals based on density functional theory in combination with the Berry-phase technique. We have chosen to investigate this property for the III-V nitrides and II-VI oxides. These are technologically significant materials and a fundamental understanding of their polarization effects is of central importance. Our results demonstrate that the trigonal structure allows a consistent determination of the spontaneous polarization of wurtzite compounds: the values obtained using this structure as a reference show an excellent agreement with the theoretical zincblende-based results reported in the literature. We have shown that the electronic and ionic part of the spontaneous polarization are of opposite sign, with the electronic term has the largest effect, indicating that the strong polarization in these materials has its origin from the partial cancellation of the two terms. Also, the spontaneous polarization was found to follow a monotonous linear behavior with the wurtzite internal parameter. Finally, we have reproduced the experimental value of the spontaneous polarization of wurtzite GaN. This was achieved by taking advantage of the stacking faults within the material. Specifically, averaging the spontaneous polarization values obtained for different types of stacking faults gives a result almost identical to the only reported experimental figure. Our theoretical approach is consistent with the experimental procedure, and can be used to determine the spontaneous polarization of other wurtzite crystals as well. We are not aware of any works where the spontaneous polarization of wurtzite stacking faults is reported. Indeed, the lack of such information was an important factor in motivating the present first-principles study.

**Acknowledgment**

The authors thank M. Texier for stimulating discussions on the subject of the present study. This work is supported by the Algerian national research projects, PRFU/DGRSDT/MESRS-Algeria (project N° B00L02UN270120220001).

**Competing interests**

The authors declare no competing interests.

**Data availability statement**

The data that support the findings of this study are available from the corresponding author upon reasonable request.

**Author contributions**

Simulation and analysis by Benbedra and Meskine. Writing by Benbedra. Supervision by Boukortt. Discussion of results and proofreading by everyone.